\documentclass[preprint,3p,12pt]{elsarticle}
\usepackage[T1]{fontenc}
\usepackage{subcaption}
\usepackage{epsfig}
\usepackage{amssymb}
\usepackage{amsthm}

\begin{document}

\begin{frontmatter}



\title{Coastlines violate the Schramm--Loewner Evolution}


\author[1]{Leidy M. L. Abril\corref{cor1}} \ead{leidy@fisica.ufc.br}

\author[2,3,4]{Erneson A. Oliveira}
  \ead{erneson@unifor.br}
  
\author[1]{André A. Moreira} \ead{auto@fisica.ufc.br}
  
\author[1]{José S. Andrade Jr.}\ead{soares@fisica.ufc.br}

\author[1,5]{Hans J. Herrmann}\ead{hans@fisica.ufc.br}

\cortext[cor1]{Corresponding author}

 \affiliation[1]{organization={Departamento de Física}, addressline={Universidade Federal do Ceará, Campus do Pici}, postcode={60451-970}, city={Fortaleza, Ceará}, country={Brazil}}

\affiliation[2]{organization={Nucleo de Ciência de Dados e Inteligência Artificial}, addressline={Universidade de Fortaleza},postcode={60811-905}, city={Fortaleza, Ceará}, country={Brazil}}

\affiliation[3]{organization={Programa de Pós Graduação em Informática Aplicada}, addressline={Universidade de Fortaleza},postcode={60811-905}, city={Fortaleza, Ceará}, country={Brazil}}

\affiliation[4]{organization={Mestrado Profissional em Ciências da Cidade}, addressline={Universidade de Fortaleza},postcode={60811-905}, city={Fortaleza, Ceará}, country={Brazil}}

\affiliation[5]{organization={PMMH}, addressline={ESPCI,7 quai St. Bernard},postcode={75005}, city={Paris}, country={France}}

\begin{abstract}
Mandelbrot's empirical observation that the coast of Britain is fractal has been confirmed by many authors, but it can be described by the Schramm--Loewner Evolution? Since the self-affine surface of our planet has a positive Hurst exponent, one would not expect a priori any critical behavior. Here, we investigate numerically the roughness and fractal dimension
of the isoheight lines of real and artificial landscapes. Using a novel algorithm to take into account overhangs, we find that the roughness exponent of isoheight lines is consistent with unity regardless of the Hurst exponent of the rough surface. Moreover, the effective fractal dimension of the iso-height lines decays linearly
with the Hurst exponent of the surface. We perform several tests to verify if the complete and accessible perimeters would follow the Schramm--Loewner Evolution and find that the left passage probability test is clearly violated, implying that coastlines violate SLE. 
\end{abstract}




\end{frontmatter}


\section{Introduction}\label{sec1}
 
  Mandelbrot\footnote{Here we consider two types of scale free objects: fractal paths which are self-similar (isotropic) and self-affine surfaces (anisotropic scaling).} established in his famous paper \textit{How long is the coast of Britain? Statistical Similarity and Fractal Dimension} \cite{mandelbrot1967long}, that coastlines are fractal objects. Since then much research has been devoted to verifying this observation. Fractal dimensions have been calculated for the coastlines of Australia \cite{mandelbrot1967long, carr1991practice,husain2021fractal}, China  \cite{xiaohua2004fractal,su2011scale,ma2016random}, the United States \cite{jiang1998fractal,carr1991practice,phillips1986spatial}, Greenland \cite{singh2013quantification},  Britain \cite{mandelbrot1967long,shelberg1983measuring} and others.
  

Coastlines are isoheight lines of Earth's topography, which is known to be self-affine~\cite{dietler1992fractal, klinkenberg1992fractal, kalda2003gradient}. This self-affine quality is prevalent across various natural phenomena, encompassing irreversible growth \cite{barabasi1995fractal, vicsek1990self}, erosional landscapes \cite{newman1990cascade}, fracture surfaces \cite{yavari2002mechanics, ponson2006anisotropic}, geologic systems \cite{carr1997statistical}, and numerous other systems.
These surfaces present long-range correlations
characterized by the Hurst exponent $H \in \{-1,1\}$. In the regime where $0<H<1$ the constructed landscape corresponds to fractional Brownian surfaces, while for $-1<H<0$ it corresponds to fractional Gaussian noise.  It was shown that, for negative H, there exists a
critical isoheight line, that  contours a critical percolating cluster,  while for positive H no criticality has ever been reported
~\cite{araujo2014recent,schmittbuhl1993percolation}. 
Since the measured Hurst exponents of natural landscapes are between 0.3 
and 0.7 \cite{dietler1992fractal,gagnon2006multifractal,valdiviezo2014hurst,crooks2018comparison}, ~\emph{i.e.}, always positive, 
the appearance of genuine critical behavior of their isoheight lines, evidenced by a well-defined fractal dimension, seems bewildering. Therefore, our aim is to investigate this issue in more detail.
  
There have been several attempts to try to model the formation 
of fractal coastlines without considering them to be just a property of a 
self-affine surface \cite{xiaohua2004fractal,sapoval2004self,luo1999formation}. On one hand, it has been argued that forces produced 
by winds, water bodies, or erosion act on several scales  \cite{husain2021fractal}. On the other hand, models have been proposed to
explain how erosion could generate stationary fractal coastlines \cite{sapoval2004self,morais2011fractality}. By studying artificially
generated landscapes, we will show that it is 
not necessary to evoke such external agents to reproduce the fractal properties
of real coastlines. Indeed, it is sufficient to evoke only the self-affinity of the Earth's surface to reveal such properties. 
 
Since surfaces with positive Hurst exponents do not exhibit criticality~\cite{araujo2014recent,schmittbuhl1993percolation}, one would naively expect that every cut through these surfaces would also be just self-affine with some roughness exponent $\alpha$.  
This exponent determines how the height fluctuations depend on the system size \cite{schmittbuhl1995reliability,bouchaud1997scaling}. Genuine self-similarity, the condition for fractality, is reached when $\alpha$ equals unity.
  We will examine the roughness exponent of real 
  coastlines   and of artificial isoheight lines extracted from self-affine
  surfaces with $H \geq 0$.  Since some coastlines, e.g.,
  in Norway, have many fjords and overhangs, which
  renders the traditional calculation of the roughness exponent very difficult,
  we were in fact compelled to develop a new algorithm to overcome such a challenge.

Critical systems are not only scale-invariant, but also exhibit conformal symmetry~\cite{saberi2013percolation}. 
Boffetta \textit{et al.}~\cite{boffetta2008winding} found that coastlines with fractal dimensions close to 4/3 pass the winding angle
test. This led them to conclude that coastlines are conformal invariant, a property 
they used to calculate the spatial distribution of diffusing particles 
reaching the coast. We will conduct more rigorous tests to evaluate the connection between coastlines and the Schramm--Loewner Evolution (SLE) theory  \cite{schramm2000scaling,cardy2005sle,bauer20062d,cardy2005sle,daryaei2012watersheds,pose2014shortest,pose2018schramm,javerzat2024schramm}.  
We shall  compare the diffusion coefficients obtained with different numerical methods for real and artificial coastlines which, in the end, will lead us to conclude that,
coastlines violate the SLE 
theory\cite{cardy2005sle,bauer20062d,daryaei2012watersheds}.

\section{Methods}
In this section, we describe the methods we employ to construct and analyze the scaling behavior of the rough surfaces and isolines. Specifically, in section~\ref{subsec:greatingsurfaces}, we review the Filtering Fourier Method and explain how we use it to generate artificial rough surfaces with the prescribed Hurst exponent. In section~\ref{subsec:methlocalw}, we present a novel method to characterize the roughness of the isoheight lines of these surfaces, as well as the real coastlines. As we will show, that independently of the Hurst exponent of the rough surface, the isoheight lines present a roughness exponent consistent with unity, suggesting that those lines are self-similar fractals rather than self-affine curves.
\subsection{Generation of self-affine surfaces\label{subsec:greatingsurfaces}}
We modeled self-affine surfaces on a square lattice by associating to each site $\textbf{r}=(x_1,x_2)$ a height $h(\textbf{r})$. Self-affine surfaces exhibit long-range correlations \cite{schmittbuhl1993percolation} and the Fourier filtering method (FFM) \cite{de2017influence,pose2018schramm,oliveira2019universal,luo2023global} is an appropriate technique to construct them.
 In the FFM, a complex-valued function is created in reciprocal space with Fourier coefficients written in the form
\begin{equation}
\eta(\textbf{q}) = \sqrt{S(\textbf{q})}u(\textbf{q})e^{2\pi \phi (\textbf{q})},
\end{equation}
where $u(\textbf{q})$ are random numbers sampled from a Gaussian distribution with mean zero and unitary variance, and the random phase $\phi(\textbf{q})$ is uniformly distributed between zero and one. The Fourier coefficients \cite{de2017influence},
\begin{equation}
c(\textbf{q})=\sqrt{S(\textbf{q})}u(\textbf{q}),
\end{equation}
 must follow the symmetry condition  $c(\textbf{-q})=\overline{c(\textbf{q})}$. Long-range correlations are imposed by considering the spectrum to be a power-law,\begin{equation}
S(\textbf{q})\sim |q|^{-\beta_c} = \left( \sqrt{q_{1}^{2}+q_{2}^{2}} \right)^{-\beta_c},
\end{equation}
where $\beta_c$ relates to the Hurst exponent through $\beta_c = 2H + d$, with the Euclidean dimension being $d=2$ for surfaces~\cite{oliveira2011optimal,barnsley1988science}.
A fast Fourier transform (FFT) of the power spectrum generates the 
real-valued surface,
\begin{equation}
h(\textbf{x}) = h(x_1,x_2) = Re\left[\sum_{q_1=0}^{N-1}\sum_{q_2=0}^{N-1} \eta(q_1,q_2) e^{-2i\pi (q_1x_1+q_2x_2)}\right],
\end{equation}
which exhibits the desired height correlations, 
$$c(\textbf{r}) =\langle (h(\textbf{x})-h(\textbf{x}+\textbf{r}))^2 \rangle \sim |\textbf{r}|^{2H},$$ where $\langle...\rangle$ represents the average over all positions $\textbf{x}$.

After creating the surfaces, isoheight lines can be generated by intersecting the surface with horizontal planes. Notably, coastlines represent isoheight lines that traverse the entire surface.

\subsection{The local width\label{subsec:methlocalw}}

\begin{figure}[b!]
     \centering
     \hfill
     \begin{subfigure}[b]{0.45\textwidth}
         \centering
         \includegraphics[width=\textwidth]{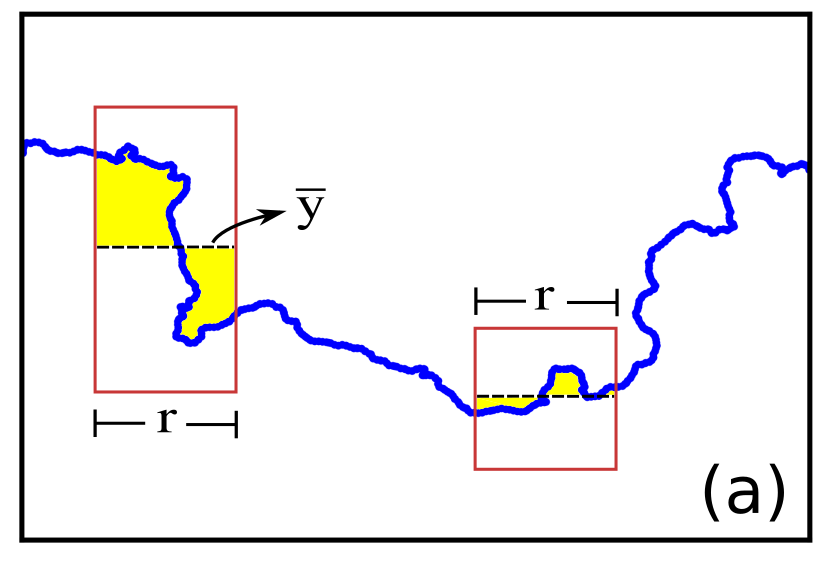}
         \label{isoheight1}
     \end{subfigure}
     \hfill
     \begin{subfigure}[b]{0.45\textwidth}
         \centering
         \includegraphics[width=\textwidth]{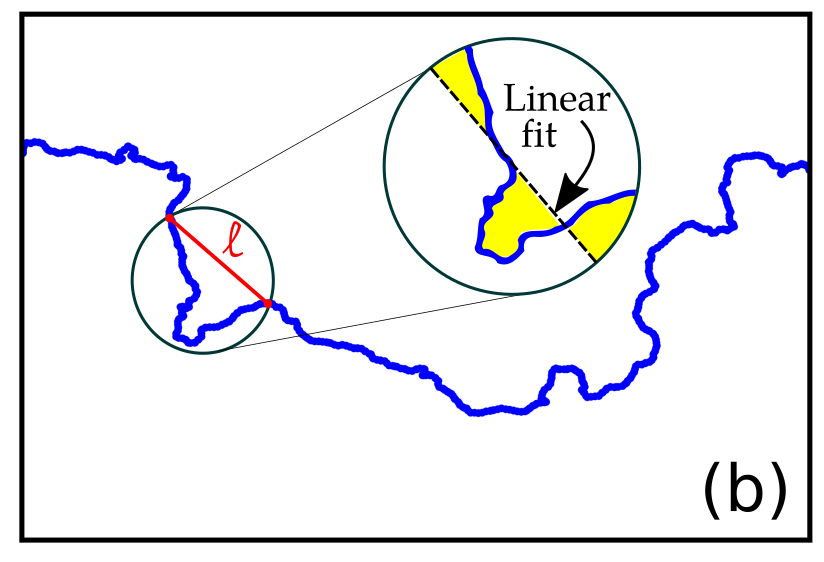}
         \label{isoheight2}
     \end{subfigure}
\caption{Schematic pictures of the two ways to implement the calculation of the width of isoheight lines: (a) Classic local width in which intervals of length $r$ are chosen in the usual way parallel to the horizontal axis and the height variances are computed with respect to the horizontal dashed lines. (b) In the modified local width, linear fits are made over sections of constant length $\ell$ along the isoheight line represented by the red segment. The fluctuations are then computed with respect to the inclined dashed line, as demonstrated in the zoomed-in view.}
\label{Fig::Isoheights}
\end{figure}

For a self-affine object, the local width scales with the length as $w(r)\sim r^\alpha $, where $\alpha$ is the roughness exponent ranging between 0 and 1. 
In the classical version \cite{family1985scaling,schmittbuhl1995reliability,krim1993roughness}, the local width of an interface describes fluctuations over a reference line corresponding to the local mean. Let's consider that the interface is described by a sequence of points in the plane $\{x_i,y_i\}$.
Fluctuations are determined by dividing the interface line in windows
of width $r$ parallel to the horizontal axis, as shown in Fig. \ref{Fig::Isoheights}-a. The fluctuation within a window
is given by the square root of the average square deviation in the $y$-axis, and the
global fluctuation $w(r)$ is the average over
all windows,
\begin{equation}
w(r) = \Bigg<\sqrt{\frac{1}{n_r}\sum_{i=1}^{n_r} \left(y_i-\bar{y}\right)^2}\,\,\Bigg>,
\label{eq:width}
\end{equation}
where $n_r$ is the number of points of the interface within a given window,
$\bar{y}$ is the average height and, the
brackets $\langle\cdots \rangle$ indicate the average over all the
 windows of length $r$ along the interface. This calculation must be performed for intervals of different lengths $r$. 
This method to calculate the width works well for interfaces 
that can be described by a single-valued function or that do not have too 
large overhangs, such as height profiles from landscapes.
Nonetheless, for some interfaces like isoheight lines which can have large overhangs and deep fjords, this procedure does not work well. For example, consider the two intervals of horizontal length $r$ shown in  Fig. \ref{Fig::Isoheights}-a,  the yellow regions -- which represent the deviations from the dashed line and thus the height fluctuations -- are not comparable for these intervals because of the different orientations of the interface. Consequently, the average width calculated using Eq. (\ref{eq:width}) exhibits very slow convergence with increasing sample size, leading to significant sample-to-sample fluctuations in the roughness exponent. Due to these reasons, we modify the traditional method to calculate the local width as follows:

\begin{itemize}
\item[(i)] Split the isoheight line in 
segments of length $\ell$, as shown in Fig. \ref{Fig::Isoheights}-b. As before we will
call these segments as windows.
\item[(ii)] For each window $k$ we 
apply a linear regression to all $n_k$  points on the isoheight line within the window, obtaining a straight line of slope $a_k$ and intercept $b_k$. 
The linear regression must be made considering the errors in both axes and for this purpose, we use an Orthogonal Distance Regression~\cite{boggs1990orthogonal}. 
\item[(iii)] Calculate the square distance between the points on the isoheight line and the straight line. The square distance from
a point $\{x_i,y_i\}$ and the line $y=a_k x +b_k$ is given by,
\begin{equation}
    d_i^2=\frac{1}{1+a^2_{k}}\Big[ y_i-a_k x_i -b_k\Big]^2.
\end{equation}
The fluctuation within a window is given
by the square-root of the the average square
distance, and the fluctuation $w(\ell)$ is 
the average over all windows,
\begin{equation}
w({\ell}) = 
\Bigg<\sqrt{\frac{1}{n_k}\sum_{i=1}^{n_k} 
d_i^2}\Bigg>.
\label{width1}
\end{equation}
\item[(iv)] Fix another distance $\ell$ and repeat steps (i) to  (iv). 
\end{itemize}


 
\subsection{ The Schramm--Loewner Evolution (SLE)}
The SLE theory encodes statistical properties of non-intersecting fractal curves through a single parameter known as the diffusion coefficient, $\kappa \in \{0,8\}$~\cite{lawler2011natural,kennedy2009numerical,zhan2019decomposition}.
These curves that in the ``chordal" representation start from the origin and grow up to infinity are conformally mapped to the upper half plane $\mathbb{H}$ with a transformation $g_t(z)$ that satisfies the Loewner differential equation, defined as \cite{gruzberg2004loewner}:
\begin{equation}
\frac{\partial g_t(z)}{\partial t} = \frac{2}{g_t(z) - \xi_t},
\label{Eq:lowewnr}
\end{equation}

\noindent where $g_0(z)=z$, $t$ is called the Loewner time and  $\xi_t$ is called the driving function. Oded Schramm found that if chordal curves 
are conformally invariant and also fulfill the Markov property then the driving function must be a one-dimensional Brownian motion  $B_t$ with a diffusion coefficient given by 
$\xi_t = \sqrt{\kappa} B_t$~\cite{schramm2000scaling}.
A necessary condition for a curve to be compatible with the SLE theory is that one must obtain the same value for the parameter $\kappa$ from four different methods \cite{daryaei2012watersheds,pose2018schramm,pose2014shortest,javerzat2024schramm}, namely,  the fractal dimension, the winding angle, the left passage probability, and the direct SLE. \\

\noindent\textbf{Fractal dimension}:  The parameter $\kappa$ is related to the fractal dimension $D_f$ through the expression~\cite{duplantier2003higher,Vicentbafarra2001},
\begin{equation}
D_f = 1+\frac{\kappa}{8}.
\label{Dfkappak}
\end{equation}
Here $D_f$ is computed using the yardstick method through $N_\epsilon\sim \epsilon^{-D_f},$ 
where $N_\epsilon$ is the number of yardsticks of size $\epsilon$ used to cover the fractal curve \cite{de2017influence}. The fractal dimension is then obtained as the slope of $N_\epsilon$ as a function of $\epsilon$, both in logarithmic scale.\\ 

\noindent\textbf{Winding angle}: The winding angle $\theta$ is 
defined as the cumulative sum of the turning angles along the curve \cite{wieland2003winding}. It starts with an initial value of $\theta_1=0$ and is computed iteratively as follows:
\begin{equation}
\theta_{i+1} = \theta_{i} + \alpha_i,
\end{equation}
where $\alpha_i$ is the angle formed between the line that connects point $i$ with point $i+1$ and the tangent line of the curve at point $i$~\cite{wieland2003winding, de2018schramm, daryaei2012watersheds}. 
For a conformally invariant curve, the variance of the winding angle increases logarithmically with size having a prefactor of $\kappa/4$~\cite{duplantier+saleur}: 

\begin{equation}
\langle\theta^{2} \rangle = b + \frac{\kappa}{4} \ln (L_y),
\label{angle}
\end{equation}
where $L_y$ is the size of the curve in the vertical direction and $b$ is a constant.\\

\noindent\textbf{Left passage probability:} 
This test is only valid for chordal curves that start at the origin and go to infinity. Therefore, one must first map original curves to the upper half plane $\mathbb{H}$ using an inverse Schwartz-Christoffel transformation~\cite{driscoll2002schwarz}. The probability that a point $z=Re^{i\phi}$ lies on the right side of the curve is given by Schramm's formula~\cite{schramm}: 

\begin{equation}
P_\kappa(\phi) = \frac{1}{2} + \frac{\Gamma\left(\frac{4}{\kappa}\right)}{\sqrt{\pi} \Gamma\left(\frac{8-\kappa}{2\kappa}\right)}\cot (\phi)\,_2F_1 \left(\frac{1}{2} ,\frac{4}{\kappa},\frac{3}{2}, -\cot^{2} (\phi)\right),
\end{equation}
where $\Gamma$ is the Gamma function and $_2F_1$ the Gauss hypergeometric function. We emphasize that the Schramm's expression for the probability only depends on $\phi$ and $\kappa$. The probability $P(\phi,R)$ that a point, with radii $R$ and angle $\phi$, given by $z=Re^{i\phi}$ is located to the right side of the curve can be computed numerically. The parameter $\kappa$ of the curve is obtained by determining the value of $\kappa$ in Schramm's formula which best fits the numerically obtained probabilities $P(\phi,R)$ for all $R$. This can be done defining the mean square deviation $Q(\kappa)$ between the two probabilities \cite{daryaei2012watersheds}:

\begin{equation}
Q(\kappa) = \frac{1}{N}\sum_R \sum_\phi \left[ P(\phi,R) - P_\kappa(\phi)\right]^2
\label{medq}
\end{equation}

and then determining the value of $\kappa$ that yields the minimum value of $Q(\kappa)$. \\

\noindent\textbf{Direct SLE}: Here, the SLE mapping is performed explicitly. This test also requires casting first the curves in chordal representation. A chordal curve is parametrized by a time $t$, called the Loewner time. In order to implement the mapping between the fractal curve and the random walk one must find the sequence of maps $g_t(z)$ from the upper half plane $\mathbb{H}$ onto itself, which satisfy the Loewner differential equation Eq.(\ref{Eq:lowewnr}). This can be numerically achieved using the vertical slit map~\cite{Kennedy},

\begin{equation}
g_t(z) = \xi_t + \sqrt{(z-\xi_t)^{2}+4\delta t},
\label{slit}
\end{equation}
where $\delta t_i=t_i-t_{i-1} =\left(Im\{z_i\} \right)^2/4$ and $\xi_{t}=Re\{z_i\}$ is the driving function.

Then the so-called zipper algorithm~\cite{Kennedy} is implemented.
Let us consider a curve that, at time $t=0$, has $N$ points $z_i(0)$ with $\xi_0=0$ and calculate how this curve $z_i(t)$ evolves when applying the mapping $g_t(z)$. After the first iteration, all the points $z_i(0)$ are mapped within the upper half plane $\mathbb{H}$ to $z_i(1) = g_1(z_{i}(0))$ with, in particular, $z_1(1)$ being now on the real axis. Then, $g_2(z)$ is applied on the remaining $N-1$ points having non-zero imaginary part. This is iteratively continued until the $i$-th point reaches the real axis following:
\begin{equation}
z_i(t) =  g_t(z_{i}(t-1)) \circ g_{t-1}(z_{i}(t-2)) \circ \cdots \circ g_1(z_{i}(0)).
\label{zipper}
\end{equation}

The parameter $\kappa$ is obtained from the variance of the driving function through the relation \cite{bauer20062d}: 
\begin{equation}
\langle \xi^{2} \rangle = \kappa t.
\end{equation}

\section{Results and Discussion}\label{results}
\begin{figure}[h!]
\centering
\begin{subfigure}{0.48\linewidth}
\includegraphics[width=\linewidth]{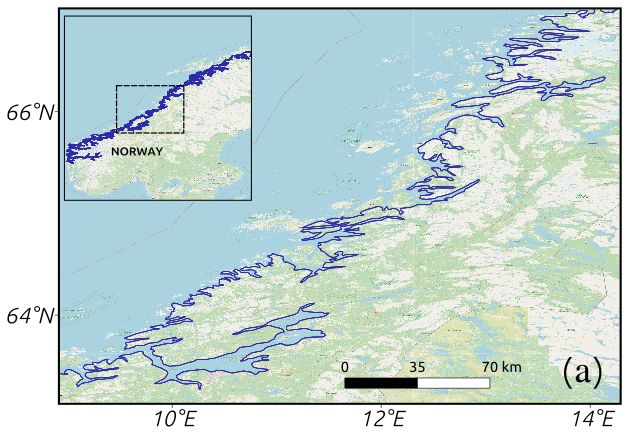} 
\label{fig:1a}
    \end{subfigure}\quad
\begin{subfigure}{0.48\linewidth}
\includegraphics[width=\linewidth]{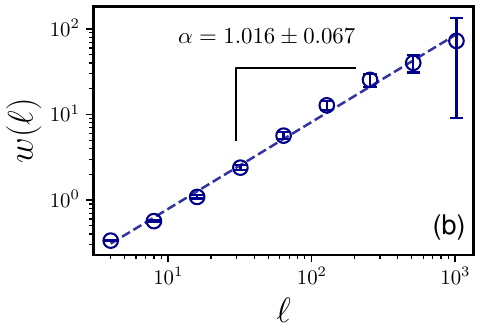}
\label{fig:1b}
    \end{subfigure}
\caption{Roughness exponent of the Norwegian coastline. (a) Section of the Norwegian coastline shown in blue. In the inset, we see the whole Norwegian coast $\sim 1300 km$ used to compute the roughness exponent, where the dashed box corresponds to the section shown in the main figure.
In (b) we show the calculation of the roughness exponent of this coastline through the log-log plot of the modified local width as a function of window size $\ell$. The roughness exponent $\alpha=1.02 \pm 0.07$ is the slope of the dashed line, which is obtained from the least-square fit to the data points of the power law, $w(\ell)\sim \ell^{\alpha}$. The bars correspond to the standard deviation. The standard errors are smaller than the symbols.}
    \label{fig:coastreales}
    \end{figure}

To begin, we analyzed the roughness exponent of both real and artificial isoheight lines. For real coastlines, we used the General Bathymetric Chart of the Oceans (GEBCO) database, which offers topographic data for the Earth's land and seafloor with a spatial resolution of $15$ arc seconds~\cite{gebco2022}. We choose ten coastlines from different regions of the Earth\footnote{The bottom-left and top-right geographical coordinates considered were
(55$^\circ$N4$^\circ$E, 71$^\circ$N23$^\circ$E) for Norway,
(4$^\circ$S48$^\circ$W, 0$^\circ$01$'$N40$^\circ$W) for Brazil,
 (7$^\circ$N77$^\circ$W, 12$^\circ$N72$^\circ$W) for Colombia,
 (41$^\circ$N13$^\circ$E, 46$^\circ$N19$^\circ$E) for Croatia, (64$^\circ$N24$^\circ$W, 66$^\circ$N15$^\circ$W) for Iceland, 
 (49$^\circ$N8$^\circ$W, 59$^\circ$N2$^\circ$E) for Britain, 
 (19$^\circ$N107$^\circ$E, 34$^\circ$N121$^\circ$E) for China,
 (32$^\circ$N78$^\circ$W, 46$^\circ$N65$^\circ$W) for USA,
 (46$^\circ$N138$^\circ$W, 60$^\circ$N122$^\circ$W) for Canada, and (18$^\circ$S122$^\circ$E, 10$^\circ$S137$^\circ$E) for Australia.
} and extracted the isoheight lines from the Digital Elevation Maps (DEM) corresponding to these regions. 
We show this procedure for the case of Norway's coastline in 
Fig.~\ref{fig:coastreales}: Plotting the width $w(\ell)$ against $\ell$ on the double-logarithmic scale, one observes a regime of constant slope.
This slope obtained with linear regression yields the roughness exponent $\alpha$.

\begin{figure}[t!]
\centering
\includegraphics[scale=1]{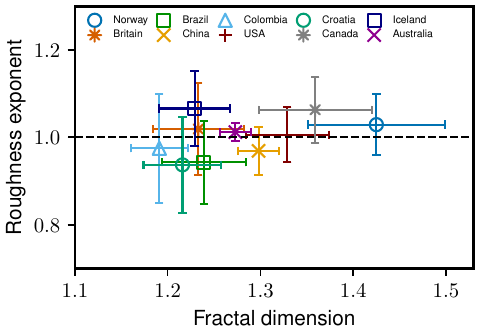}
\caption{Roughness exponent against the fractal dimension for ten different coastlines. The bars correspond to the standard error of the regression. The black dashed line represents the value of the roughness exponent one expects for a fractal curve. 
}
\label{Fig::alphavsDf}
\end{figure}

We estimated the roughness exponents and the fractal dimensions for the real coastlines, as shown in Fig.~\ref{Fig::alphavsDf}. 
We see that, while the fractal dimensions $D_f$ of the coastlines vary significantly, the roughness exponents $\alpha$  with their standard deviations remain
always around unity. A roughness exponent $\alpha \approx 1$ is a clear indication for isotropic scale-free behavior and thus for genuine fractality \cite{schmittbuhl1995reliability}.

\begin{figure}[b!]
\centering
    \begin{subfigure}{0.31\linewidth}
\includegraphics[width=\linewidth]{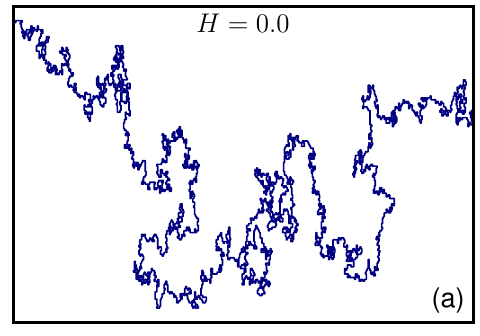} 
    \end{subfigure}\quad
    \begin{subfigure}{0.31\linewidth}
\includegraphics[width=\linewidth]{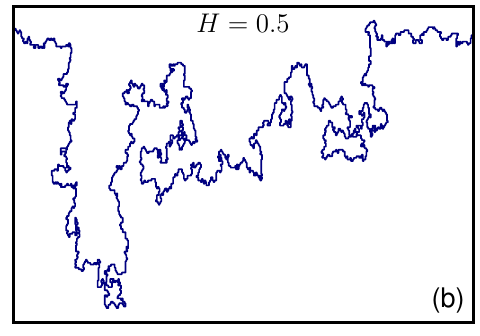}
    \end{subfigure}
    \quad
    \begin{subfigure}{0.31\linewidth}
\includegraphics[width=\linewidth]{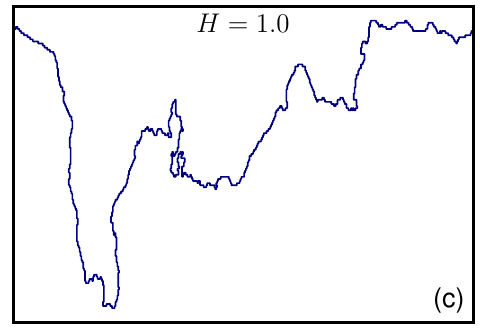}
   \end{subfigure}
      \caption{Complete perimeters for artificial surfaces with different Hurst exponents. (a) corresponds to the Gaussian free field and is very rough presenting many overhangs, (b) shows the Brownian motion showing less irregularity, and (c) displays a smoother line where overhangs tend to disappear.}
   \label{fig:lineasartificialesa}
   \end{figure}

\begin{figure}[h!]
\centering
\includegraphics[scale=0.89]{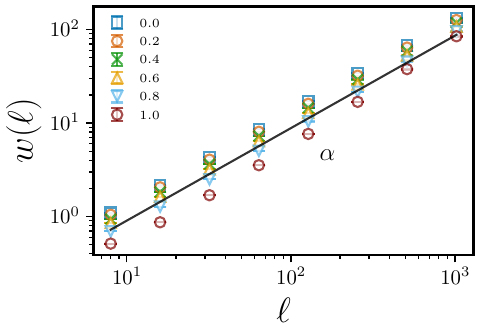}
\caption{Modified local width $w(\ell)$ of complete perimeters as a function of  segment size $\ell$ for surfaces with $H\geq0$.  The roughness exponent $\alpha$ corresponds to the slope of the straight line drawn in the log-log scale.  The bars correspond to the variance over $10^3$ samples.}
\label{Fig::width}
\end{figure}

\begin{figure}[b!] 
\centering
\includegraphics[scale=0.89]{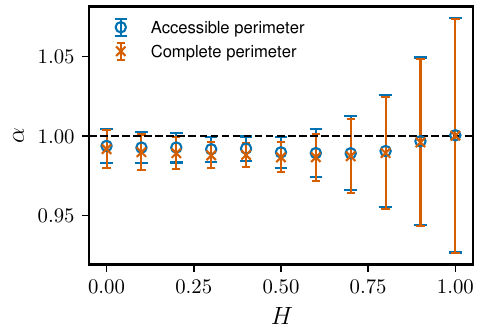}
\caption{Roughness exponents $\alpha$ for the complete and accessible perimeters from self-affine surfaces with a linear size $L=4096$, plotted against the Hurst exponent of the surface.
 The bars correspond to the standard deviation calculated from 10$^3$ samples for each Hurst exponent. The standard errors are smaller than the symbols. 
 The roughness exponent of coastlines is consistent with $\alpha\approx1$ for all Hurst exponents of surfaces, which indicates that the coastlines are fractal. 
}
\label{Fig::R_ExponentvsH}
\end{figure}

To create artificial coastlines, we generated surfaces of size  $4096\times 4096$ with Hurst exponents in the range $0\leq H\leq1$. We then proceeded to extract the isoheight lines of each surface.  For every isoheight line that spans the entire surface, we observe two distinct paths: the complete and accessible perimeters. 
The complete perimeter corresponds to all the bonds along the isoheight line \cite{de2017influence}. The accessible perimeter is a shorter path derived from the complete perimeter in which bridges of size $\epsilon$ are used to bypass overhangs and fjords. This is implemented within the yardstick method when going along the curve from point $i$ to $j$. Specifically, one draws a circle of radius $\epsilon$ around point $i$. By selecting the first crossing of the circle with the curve as point $j$, one obtains the complete perimeter. Conversely, choosing the last crossing of the circle with the curve as point $j$ yields the accessible perimeter. To complete the yardstick method one must do this for all points $i$ along the curve to obtain the total number of yardsticks. The procedure is then repeated for different values of $\epsilon$ to extract the fractal dimension~\cite{schrenk2013percolation}.

The complexity of an isoheight line diminishes the more the Hurst exponent associated with the surface increases. This can be seen in Fig.  \ref{fig:lineasartificialesa} for the case of the complete perimeter. While for $H=0$ there are numerous overhangs present, they virtually disappear for $H=1$.  

\begin{figure}[h!]
\centering
\includegraphics[scale=0.90]{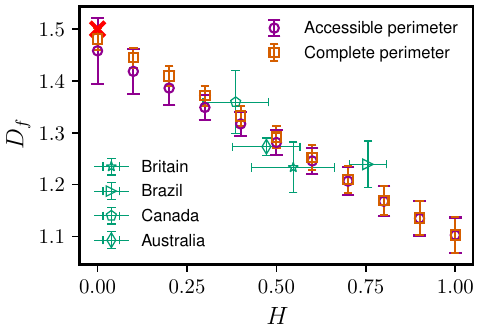}
\caption{Fractal dimension of artificial and real coastlines as a function of the Hurst exponent of the surface. For the artificial coastlines, we use the complete and accessible perimeters from  $10^{3}$ surfaces with different Hurst exponents. 
The red cross indicates the exactly known result $D_f = 3/2$ for the Gaussian free field, $H=0$.
For real coastlines, we compute both the fractal dimension of the complete perimeter and the Hurst exponent of each surface. The bars represent the standard deviations. The standard errors are smaller than the symbols.}
\label{Fig::DvsH}
\end{figure}

We calculated the local width for both complete and accessible perimeters from surfaces with different Hurst exponents. The modified local width for the complete perimeter as function of distance $\ell$ is depicted in Fig. \ref{Fig::width}.
Notably, we observe a linear relationship when viewed on a double logarithmic scale, spanning three orders of magnitude.
In Fig. \ref{Fig::R_ExponentvsH} we show roughness exponents for the complete and accessible perimeters for surfaces with different Hurst exponents $H\geq0$. 
We observe that the values of the roughness exponents are statistically consistent with unity for all Hurst exponents, which is a clear indication that the isoheight lines are fractal, despite having been extracted from self-affine surfaces. This result is another confirmation that coastlines are genuine fractals as already claimed by Mandelbrot \cite{mandelbrot1967long}.

Using the yardstick method, we calculated the fractal dimension of both the complete and accessible perimeters for surfaces with different Hurst exponents, as shown in Fig. \ref{Fig::DvsH}. Our results are consistent with the values obtained in Refs.~\cite{pose2018schramm} and \cite{de2017influence}. For the  Gaussian free field \cite{lodhia2016fractional}, when the surface has $H=0$, the fractal dimension of both paths is exactly known to be $D_f = 3/2$ (see red cross in Fig.~\ref{Fig::DvsH}) \cite{de2017influence,pose2018schramm, kondev1995geometrical}.

In the following, we present numerical tests of the SLE theory applied to the coastlines. Our procedure consists in calculating the parameter $\kappa$ with different numerical methods and compare the obtained values with each other. 
If coastlines are compatible with SLE, the values of $\kappa$ obtained through different numerical methods should coincide \cite{pose2018schramm,daryaei2012watersheds,de2018schramm}. 

\begin{figure}[t!]
\centering
\includegraphics[scale=1]{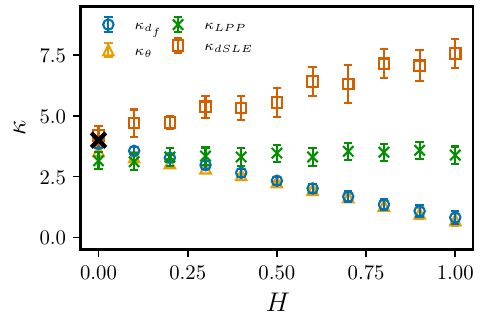}
\caption{Diffusion coefficients obtained with four different methods for the complete perimeter of isoheight lines from self-affine surfaces with different Hurst exponent. The black cross corresponds to the analytical result for $H=0$. 
We use 300 samples for each Hurst exponent and the bars correspond to the standard deviations. The standard errors are smaller than the symbols. 
}
\label{Fig::kappavsH}
\end{figure}

We computed the diffusion coefficient $\kappa$ for the complete perimeter of both artificial isoheight lines and real coastlines through the fractal dimension ($\kappa_{D_f}$), the winding angle ($\kappa_\theta$), the left passage probability ($\kappa_{LPP}$) and direct SLE ($\kappa_{dSLE}$).
Since we had already calculated the fractal dimension, we can just obtain $\kappa_{D_f}$ applying Eq.~(\ref{Dfkappak}). In Fig.~\ref{Fig::kappavsH} and Table~\ref{Tab:realcoasts}, we show $\kappa_{D_f}$ for different Hurst exponents for artificial and real isoheight lines, respectively.

\begin{figure}[b!]
\centering
\begin{subfigure}{0.48\linewidth}
\includegraphics[width=\linewidth]{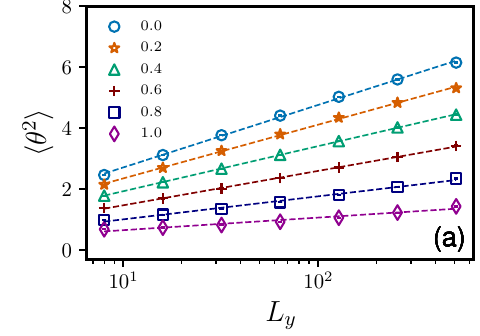} 
\end{subfigure}\quad
\begin{subfigure}{0.48\linewidth}
\includegraphics[width=\linewidth]{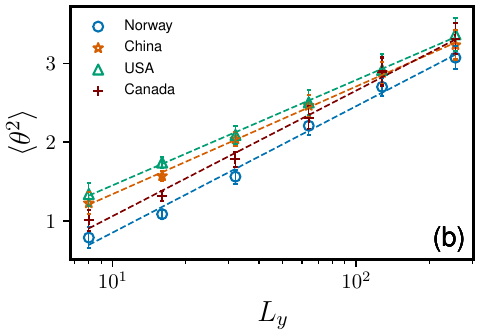}
\end{subfigure}
\caption{Winding angle statistics of isoheight lines. (a) Variance of the winding angle as a function of the vertical size of artificial complete perimeters for different Hurst exponents. The bars correspond to the standard deviation of 300  samples. 
(b) Variance of the winding angle for four natural coastlines.  We compute the variance of the angle in sections of length $L_y$ covering the coastline and then average over sections. The bars represent the standard deviations of values obtained for each length $L_y$.}
\label{Fig:windingangle}
\end{figure}

In order to calculate the parameter $\kappa_\theta$ through the winding angle one starts on the isoheight line at the bottom edge of the surface. The winding angle is then accumulated along the isoheight line and its variance monitored as a function of the vertical distance $L_y$.  
Figure~\ref{Fig:windingangle} shows the variance $\langle \theta^{2} \rangle$
of the winding angle for both artificial and real isoheight lines as a function of the logarithm of $L_{y}$ and for different values of the Hurst exponent.  Considering the consistent linear behavior of all curves, the values of $\kappa_{\theta}$ are then obtained from the least-squares fits to the data points, as per Eq.~(\ref{angle}), based on their slopes. In Fig.~\ref{Fig::kappavsH} and Table~\ref{Tab:realcoasts}, we also show $\kappa_{\theta}$ for different Hurst exponents for artificial and real isoheight lines respectively.

\begin{figure}[h!]
\centering
    \begin{subfigure}{0.48\linewidth}
\includegraphics[width=\linewidth]{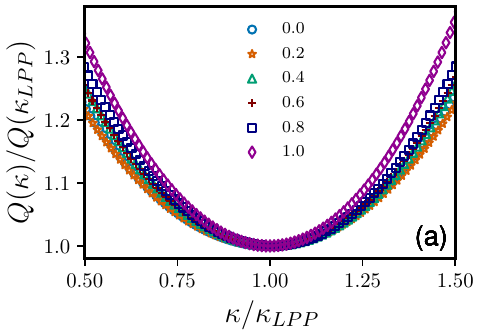} 
\label{fig:a2}
    \end{subfigure}\quad
    \begin{subfigure}{0.48\linewidth}
\includegraphics[width=\linewidth]{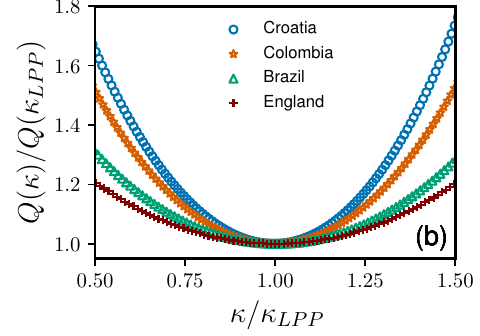}
\label{fig:03}
\end{subfigure}
\caption{Left passage probability for coastlines. In (a) the rescaled mean square deviation $Q(\kappa)/Q(\kappa_{LPP})$ is shown as a function of $\kappa/\kappa_{LPP}$ for artificial isoheight lines and various Hurst exponents. We have analyzed 300 artificial curves in the range $L_y \in [0,128]$ using 256 points in the upper half plane $\mathbb{H}$ inside the interval $0< R \leq 1$, and with angles $\phi$ in steps of $\pi/150$ within $0\leq \phi \leq \pi$. For each curve, we determine the position of the minimum $\kappa_{LPP}$ and then shift all curves on top of each other by dividing the horizontal axis by $\kappa_{LPP}$.  
(b) The rescaled mean square deviation $Q(\kappa)/Q(\kappa_{LPP})$ against $\kappa/\kappa_{LPP}$ for four real coastlines. We have analyzed about $5 \times 10^5$ points for different radii $R$ and angles $\phi$. As in (a), all curves are shifted as to fall on each other by dividing the horizontal axis by the position of the minimum $\kappa_{LPP}$.}
\label{Fig:leftpp}
\end{figure}

For the left passage probability and the direct SLE tests, we must make a Schwartz-Christoffel transformation~\cite{driscoll2002schwarz} of the isoheight lines in order to obtain the chordal representation of the curves.
To do so, we need to restrict the curve to a square and then map its vertices to the real axis of the upper half plane $\mathbb{H}$. We chose as square the domain $[-1,1]\times [0,2]$  and restricted the isoheight line to be inside the domain $[-0.5,0.5]\times [0,1]$ \cite{daryaei2012watersheds}. Then, we mapped conformally the curve to the upper half plane $\mathbb{H}$. Once in the chordal representation, we can apply on the curve the left passage probability test, namely, we calculate numerically the mean square deviation $Q(\kappa)$ given by Eq.~(\ref{medq}) and then identify the value $\kappa_{LPP}$ for which $Q(\kappa)$ is minimized, as shown in Fig.~\ref{Fig:leftpp}. The values of $\kappa_{LPP}$ are shown in Fig.~\ref{Fig::kappavsH} and Table~\ref{Tab:realcoasts} as a function of Hurst exponents for artificial and real isoheight lines, respectively. 

\begin{figure}[b!]
\centering
    \begin{subfigure}{0.48\linewidth}
\includegraphics[width=\linewidth]{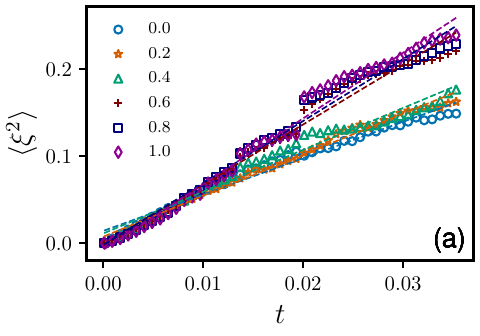} 
\label{fig:artdsle}
    \end{subfigure}\quad
\begin{subfigure}{0.48\linewidth}
\includegraphics[width=\linewidth]{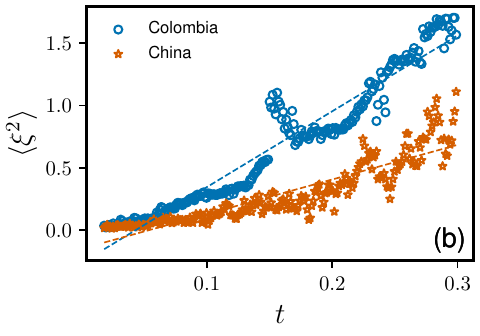}
\label{fig:realdsle}
\end{subfigure}
\caption{Direct SLE. Mean square displacement of the driving function as function of time for (a) 300 artificial isoheight lines for different Hurst exponents and (b) real coastlines of Colombia and China.
For the real coastlines, the linear behavior is not evident.}
\label{Fig:dsle}
\end{figure}
Finally, to implement the direct SLE, we took each isoheight line mapped in the upper plane and applied on it the zipper algorithm in the vertical discretization~\cite{Kennedy}, given by Eqs.~(\ref{slit}) and (\ref{zipper}). The diffusion parameter $\kappa_{dSLE}$ corresponds to the slope of the mean square displacement $\langle\xi^2\rangle$ of the driving function as a 
 function of time $t$, as shown in Fig.~\ref{Fig:dsle}. 
 For certain real coastlines, such as those of Colombia and China, the relationship of  $\langle\xi^2\rangle$ versus $t$ exhibits a linear behavior making the determination of $\kappa_{dSLE}$ difficult.

\begin{table}[t!]

\begin{center}

\renewcommand{\arraystretch}{1.2}
\begin{tabular}{l|c|c|c|c|c}
Coastline & $H$& $\kappa_{D_f}$ & $\kappa_{\theta}$ & $\kappa_{LPP}$ & $\kappa_{dSLE}$ \\ 
\hline 
Norway & $0.395 \pm 0.083$& $3.400 \pm0.592$ & $ 2.792\pm  0.094$ & $ 4.412\pm0.067$& $7.893 \pm 0.034$\\ 

Britain &$0.547\pm 0.369$ & $1.864 \pm 0.392$ & $1.816 \pm 0.085$ & $3.040 \pm0.067$ & $5.757 \pm 0.159$\\ 

Brazil & $0.756 \pm 0.051$& $1.912 \pm 0.360$ &$ 1.858\pm0.170$ & $ 3.146\pm0.067$ & $3.988\pm 0.021$ \\ 

China &$0.411\pm 0.154$ & $2.384\pm 0.176$ & $2.377 \pm 0.034$ & $1.704 \pm 0.067$&$2.635 \pm 0.015$\\ 

Colombia&$0.586 \pm 0.086 $  & $1.528\pm 0.248$& $1.490 \pm0.082$ & $3.603 \pm 0.067$ & $5.862 \pm 0.018$ \\ 

USA &$0.640\pm 0.174$ & $2.632\pm 0.360$ & $ 2.327\pm 0.030$ & $4.307 \pm 0.067$& $8.333\pm 0.030$\\ 

Croatia& $0.571\pm 0.235$ & $1.728\pm 0.336$ & $1.708 \pm 0.111$& $ 4.025\pm0.067$ &$6.700\pm 0.023$\\ 

Canada &$0.386 \pm 0.091$ & $2.872 \pm 0.488$& $ 2.772\pm0.081$ & $3.075 \pm0.067$ &$6.026 \pm 0.029$\\

Iceland &$0.590 \pm 0.276$ & $ 1.832\pm0.304$ & $1.818 \pm 0.088$ & $3.970 \pm0.067$ & $6.761\pm 0.153$\\ 

Australia & $0.472 \pm 0.094$& $2.184 \pm0.136$ & $ 2.004\pm0.050$ & $4.060 \pm 0.067$&$4.258\pm 0.040$\\ 

\end{tabular} 
\end{center}
\caption{Results of the SLE theory applied to the coastlines. 
The Hurst exponent is computed for each surface where the corresponding coastline is situated, while the diffusion coefficients are determined directly from the coastline itself.
The values for $\kappa_{d_f}$ and $\kappa_{\theta}$ are very similar indicating that conformal invariance might hold. However, $\kappa_{LPP}$ and $\kappa_{dSLE}$ deviate significantly from the other $\kappa$ values, showing that coastlines are not compatible with the SLE theory.} 
\label{Tab:realcoasts}
\end{table}

The results for the four tests applied to artificial coastlines are summarised in Fig. \ref{Fig::kappavsH}. It is evident that the $\kappa$ parameters evaluated by different tests do not yield consistent values. From this observation, we draw the conclusion that the coastlines do not conform to the Schramm--Loewner theory. 
This conclusion is consistent with the findings of Posé \textit{et. al}~\cite{pose2018schramm}, who reported that the accessible perimeter of artificial isoheight lines from surfaces with
$H>0$ is not compatible with SLE theory, in contrast to their findings $H\leqslant0$~\cite{de2018schramm}. 

In Table~\ref{Tab:realcoasts}, we present the values for the diffusion parameter obtained for real coastlines from four tests. One can observe that $\kappa_{D_f}$ and $\kappa_{\theta}$  have similar values, unlike $\kappa_{LPP}$ and $\kappa_{dSLE}$ which are quite far apart. The most surprising discrepancy emerged from the analysis of the coastline of China, where $\kappa_{LPP}$ and $\kappa_{dSLE}$ are very different from what we would expect from our simulations shown in Fig.~\ref{Fig::kappavsH}. 

\section{Conclusions}
We have studied the scaling properties of real coastlines and isoheight lines of self-affine surfaces. To analyze the roughness of these curves, which exhibit numerous overhangs, we introduce a novel method to compute the local width.  We show that, independent of the Hurst exponents of the surface, the roughness exponents of the isoheight curves are always unity, consistent with a self-similar isotropic fractal geometry for these curves.


 On the other hand, we have applied the SLE theory to the artificial and real isoheight lines by computing the diffusion parameter through four distinct numerical tests. 
 We have shown that the values of the diffusion parameter do not agree with each other and, therefore, concluding that coastlines are not compatible with the SLE theory. 
 
\section*{Acknowledgements}
We thank Nuno AM Araújo for useful discussions. 
We thank the Brazilian agencies CNPq, CAPES, FUNCAP, the National Institute of Science and Technology for Complex Systems (INCT-SC) in Brazil, and the Edson Queiroz Foundation for financial support.



  \bibliographystyle{elsarticle-num} 
  \bibliography{sn-bibliography}


\end{document}